# Demonstration of distinct semiconducting transport characteristics of monolayer graphene functionalized via plasma activation of substrate surfaces

*Po-Hsiang Wang[1], Fu-Yu Shih[1,2], Shao-Yu Chen[1], Alvin B. Hernandez[1,2], Po-Hsun Ho[3], Lo-Yueh Chang[4,5], Chia-Hao Chen[4], Hsiang-Chih Chiu[6], Chun-Wei Chen[3], and Wei-Hua Wang[\*,1]*

[1]Institute of Atomic and Molecular Sciences, Academia Sinica, No. 1, Roosevelt Rd., Sec. 4, Taipei 10617, Taiwan

[2]Department of Physics, National Taiwan University, No. 1, Roosevelt Rd., Sec. 4, Taipei 10617, Taiwan

[3]Department of Materials Science and Engineering, National Taiwan University, No. 1, Roosevelt Rd., Sec. 4, Taipei 10617, Taiwan

[4]National Synchrotron Radiation Research Center, 101 Hsin-Ann Road, Hsinchu 30076, Taiwan

[5]Department of Physics, National Tsing-Hua University, 101 Kuang-Fu Rd., Sec. 2, Hsinchu 30013, Taiwan

[6]Department of Physics, National Taiwan Normal University, 162 Heping East Rd., Sec. 1, Taipei 10610, Taiwan

[\*]Corresponding Author. Tel: +886-2-2366-8208, E-mail: wwang@sinica.edu.tw (W.-H. Wang)

**Abstract**

We report semiconducting behavior of monolayer graphene enabled through plasma activation of substrate surfaces. The graphene devices are fabricated by mechanical exfoliation onto pre-processed $SiO_2$/Si substrates. Contrary to pristine graphene, these graphene samples exhibit a transport gap as well as nonlinear transfer characteristics, a large on/off ratio of 600 at cryogenic temperatures, and an insulating-like temperature dependence. Raman spectroscopic characterization shows evidence of $sp^3$ hybridization of C atoms in the samples of graphene on activated $SiO_2$/Si substrates. We analyze the hopping transport at low temperatures, and weak localization observed from magnetotransport measurements, suggesting a correlation between carrier localization and the $sp^3$-type defects in the functionalized graphene. The present study demonstrates the functionalization of graphene using a novel substrate surface-activation method for future graphene-based applications.

**Manuscript text**

**1. Introduction**

Many interesting electrical properties of graphene[1-3] have been demonstrated and are attributed to its massless electronic structure.[4, 5] However, pristine graphene is a zero-gap material with finite conductivity at the Dirac point, which limits its potential for electronic applications.[6] To date, several approaches have been explored to induce an energy gap in graphene including functionalization,[7-10] quantum confinement[11-13], and substrate-induced lattice mismatch.[14, 15] Breaking of the inversion symmetry in bilayer graphene has been demonstrated with the application of a perpendicular electric field[16-19] or surface adsorbates.[20-22] Moreover, a disorder-induced transport gap was introduced in graphene samples exhibiting structural disorder at the edge[23-25] or in the bulk.[26]

Alternatively, an energy gap can be created in monolayer graphene through the substrate effect. This approach is very attractive because the band-gap area of graphene can be specified and controlled, and therefore, both transistors and interconnects can be carved out from a single sheet of graphene. Examples of this approach include breaking the sublattice symmetry in epitaxial graphene[27] and graphene/boron nitride stacking.[28] Additionally, functionalization of graphene by controlling the surface reactivity of the substrate and subsequently modifying graphene has been reported.[29, 30] However, functionalization of graphene has not been demonstrated previously by engineering the chemical activation of the substrate surface, followed by simple exfoliation of graphene on substrates without further treatment on graphene. Moreover, establishment of noticeable semiconducting transport characteristics in graphene using the substrate effect has been lacking. Here, we present experimental observations of the functionalization of graphene induced by chemically activated substrate surfaces (Figure 1a). We show extensive transport characteristics in monolayer graphene samples including a transport gap, nonlinear transfer characteristics, and a large on/off ratio of 600 at cryogenic temperatures. Detailed analyses of Raman spectroscopy characterization, hopping transport behaviors, and magnetotransport properties of the graphene devices indicate consistent functionalization of graphene.

## 2. Device Fabrication

The devices of graphene on activated $SiO_2$/Si were fabricated using conventional $SiO_2$/Si substrates. The detailed procedure of the device fabrication can be found in the Supplementary Data S1. Briefly, the surface of the $SiO_2$/Si substrate was treated with oxygen plasma for 10 min to increase the density of silanol groups on the surface.[31] The substrates were then dipped in water to assist the silanol group formation,[32] followed by blow-drying with $N_2$. After treatment, the $SiO_2$/Si substrates exhibited much lower contact angles (< 20 degrees) compared to untreated substrates, indicating high surface hydrophilicity attributed to the presence of activated polar groups.[30] Monolayer graphene was then mechanically exfoliated onto the activated $SiO_2$/Si substrates. A TEM grid was used as a shadow mask to define the electrical contact areas, and Ti/Au (5 nm/50 nm) were deposited as the electrical contacts. The resist-free fabrication method was employed to avoid resist residue, which would lead to undesirable effects on the transport properties. Unless otherwise specified, all electrical measurements were performed using standard lock-in techniques with an AC bias current of less than 10 nA.

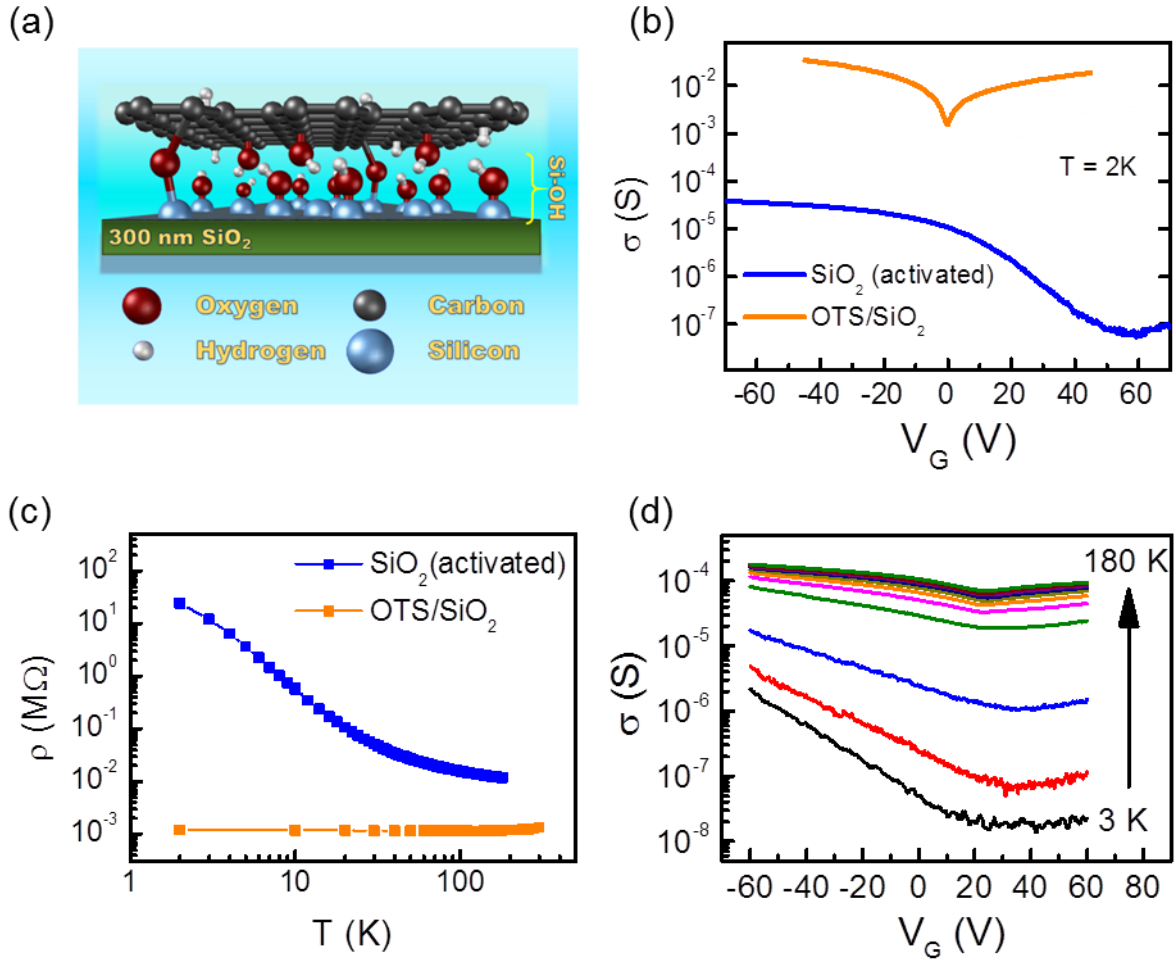

**Figure 1. Device Characteristics** (a) A Schematic diagram of the functionalized graphene on an activated SiO$_2$/Si substrate (not to scale). (b) Comparison of $\sigma - V_G$ curves between a sample of graphene on an activated SiO$_2$/Si substrate (blue curve, sample A) and a sample of graphene on OTS-modified SiO$_2$/Si substrate (orange curve, control sample). (c) Comparison of the temperature dependence of $\rho_{CNP}$ between sample A (blue squares) and the control sample (orange squares). (d) The temperature dependence of the $\sigma - V_G$ curves for sample B ranging from 3 to 180 K.

## 3. Results and Discussion

### 3.1 Temperature dependence of the transport properties.

We first compare the conductivity versus gate voltage ($\sigma - V_G$) curves (Figure 1b) of graphene on an activated SiO$_2$/Si substrate (sample A) and a control sample consisting of graphene on an octadecyltrichlorosilane (OTS)-functionalized SiO$_2$/Si substrate to demonstrate how the unique substrate surface treatment affects the electronic properties of graphene.[33] The control sample showed transport properties resembling that of intrinsic graphene including high mobility (~ 60,000 cm$^2$/Vs), a low on/off ratio (~ 10), and small residual doping, which can be attributed to very small interaction between graphene and the OTS-functionalized substrates. In contrast, sample A exhibited a significant reduction in device conductivity and mobility ($\mu_h$ ~ 50 cm$^2$/Vs), and a large on/off ratio of 600. Figure 1c shows the temperature ($T$) dependence of the channel resistivity at the charge neutrality point ($\rho_{CNP}$) for sample A and the control sample. The $\rho_{CNP}$ of the control sample showed negligible $T$ dependence, because pristine graphene lacks a band gap and exhibits weak electron-phonon scattering. Conversely, sample A exhibited insulating behavior observed by $\rho_{CNP}$ increasing more than three orders of magnitude (from 0.01 M$\Omega$ at 200 K to 20 M$\Omega$ at 2 K). The distinct differences in transport properties between sample A and the control sample indicate that the graphene/substrate interaction can effectively alter the transport properties of pristine graphene.

Figure 1d shows the $\sigma - V_G$ curves at different $T$ of another device made with graphene on an activated SiO$_2$/Si substrate (sample B). The $\sigma - V_G$ curves of sample B exhibited strong $T$ dependencies, unlike those of pristine graphene; they are shown in logarithmic scale to reveal the large changes in $\sigma$ at lower $T$. Moreover, the on/off ratio increases from 2 at 180 K to 200 at 3 K. In addition to the strong $T$ dependence of $\sigma$ at the charge neutrality point (CNP) presented earlier, the $\sigma$ at high carrier density ($V_G = -60$ V) decreases by approximately 2 orders of magnitude, revealing an entirely different behavior compared to pristine graphene. The high resistivity and enhanced on/off ratio at low $T$ strongly suggests the formation of an energy gap in the samples of graphene on activated SiO$_2$/Si substrates.

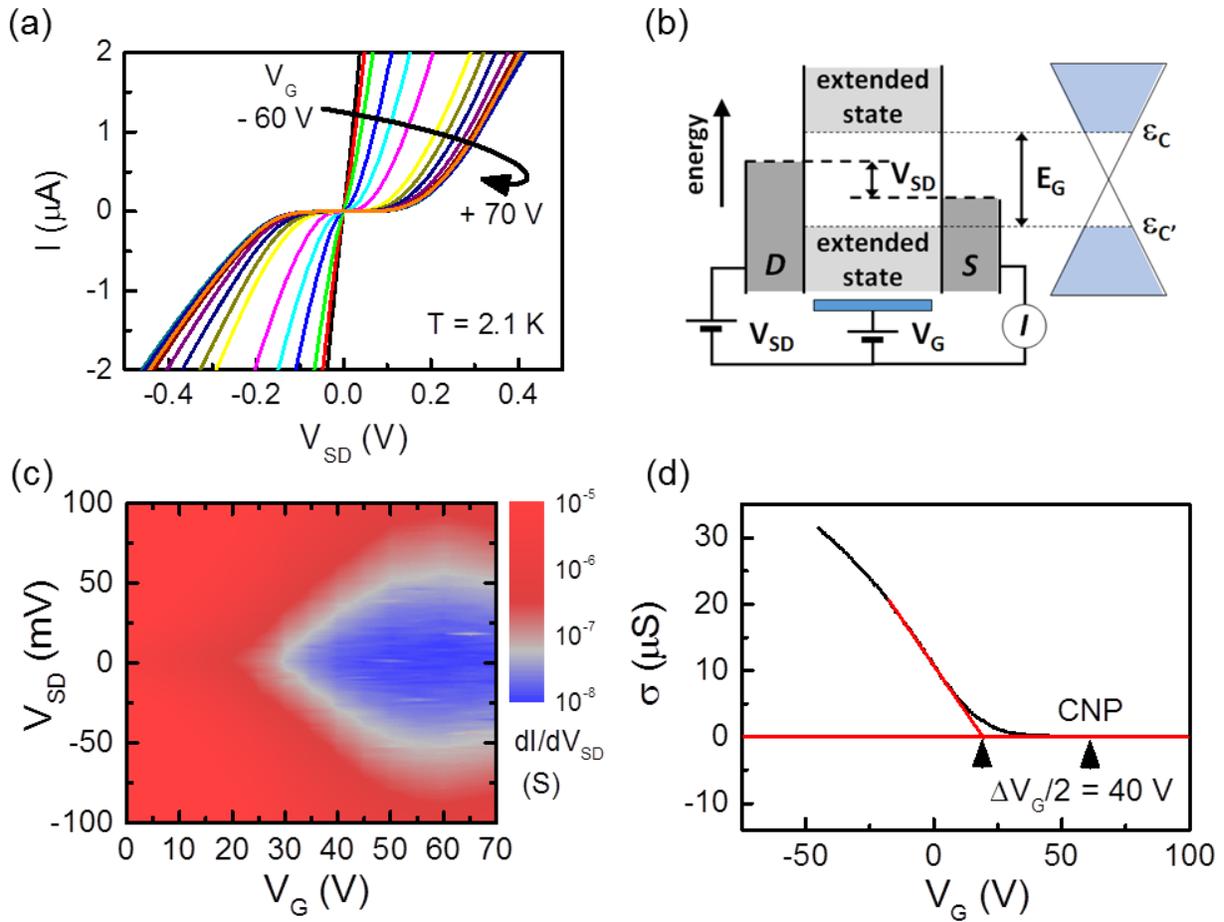

**Figure 2. Energy gap estimation of graphene on activated SiO$_2$/Si devices** (a) Nonlinear $I-V_{SD}$ curves at different $V_G$ for sample A measured at $T = 2$ K. The $I-V_{SD}$ curve corresponding to the CNP ($V_G = 60$ V) shows the most pronounced nonlinear characteristics. $V_G$ ranges from -60 V to 70 V, as shown by the curved arrow. (b) Schematic energy diagram of a device with $E_G$ under an applied gate voltage $V_G$ and bias $V_{SD}$, where $\varepsilon_C$ and $\varepsilon_{C'}$ are conduction

and valence band edges, respectively. (c) The $dI/dV_{SD}$ mapping as a function of $V_G$ and $V_{SD}$ measured at $T = 2.1$ K. The blue area indicates the turned-off region and the vertical indices of the diamond shape determine the energy gap. (d) The evaluation of the transport gap, $\Delta V_G$, as measured for $V_G$ at $T = 2.1$ K. $\Delta V_G$ is estimated as twice the $V_G$ difference between the CNP and the intersection of red lines (21.2 – 41 V, designated by arrows).

### 3.2 Transport gap at cryogenic temperature.

We show further evidence of the transport gap in the sample of monolayer graphene on activated SiO$_2$/Si substrate at low $T$. Figure 2a displays DC measurements of current vs. source-drain voltage ($I - V_{SD}$) curves at different $V_G$ for sample A at $T = 2$ K. The $I - V_{SD}$ curves show that the most pronounced nonlinearity occurs at $V_G = 60$ V, which corresponds to the CNP of the sample (Figure 1b, blue curve). The observation of a low conduction regime in the nonlinear $I - V_{SD}$ curve suggests the presence of an energy gap ($E_G$) acting as a potential barrier for the carriers. At high carrier density, the $I - V_{SD}$ curve gradually becomes more linear, indicating that the graphene transforms to exhibit metallic behaviors. This transition is consistent with the notion of energy gap formation in the graphene samples. By adjusting $V_G$, the Fermi level shifts from within the energy gap to the extended states, leading to a change in the carrier transport behavior from insulating to metallic in nature.

We now estimate the size of the transport gap by examining the differential conductance ($dI/dV_{SD}$) in the nonlinear regime as a function of $V_G$ and $V_{SD}$.[34] Figure 2b shows a schematic energy diagram of a graphene device with $E_G$, along with source, drain, and back-gate electrodes. The source and drain levels are varied with applied $V_{SD}$, and the position of $E_G$ relative to the source-drain energy levels is controlled by $V_G$. When the mobility edges overlap with the bias window between the source and drain levels, the channel becomes conducting and the current rises markedly. Figure 2c shows $dI/dV_{SD}$ versus $V_G$ and $V_{SD}$ for sample A at $T = 2$ K. The blue area represents the turned-off region in the $V_G - V_{SD}$ plane, which is diamond-shaped, indicating that both $V_G$ and $V_{SD}$ influence the position of the mobility edges relative to the source and drain energy levels. We can then obtain $E_G \sim 100$ meV from the value of $V_{SD}$ at the vertices of the diamond-shaped area. We note that semiconducting transport characteristics are also found in other samples of graphene on activated $SiO_2$/Si substrates, which show comparable transport gaps in the range of ~ 80 – 100 meV (Supporting Information S2).

We also estimate the energy in the single particle energy spectrum ($\Delta_m$) based on the transport gap region, as measured in $V_G$. First, we obtain the transport gap $\Delta V_G / 2 = 40$ V from the $\sigma - V_G$ curve of sample A,[35] as shown in Figure 2c (here we assume that the conductivity turn-on is symmetric to the CNP; turn-on of electron conduction was not possible due to the risk of high applied $V_G$). We can then estimate the $\Delta_m$ corresponding to the transport gap from $\Delta_m \approx \hbar v_F \sqrt{2\pi C_G \Delta V_G / |e|}$, where $v_F = 10^6$ m/s is the Fermi velocity of graphene and $C_G = 115$ aF/$\mu$m$^2$ is the back-gate capacitance per unit area.[24] We obtained $\Delta_m \approx 400$ meV for sample A, which is larger than the $E_G$ derived from Figure 2b. This discrepancy has previously been observed in disordered graphene and is attributed to different physical meanings of these two energy scales.[23, 24]

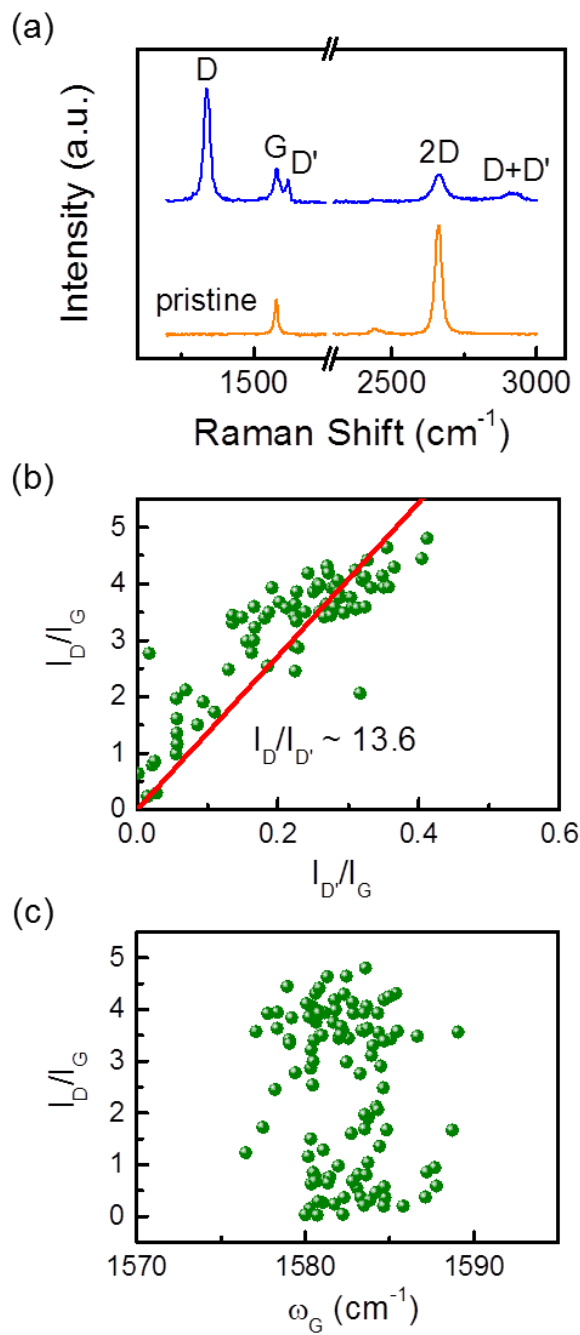

**Figure 3. Raman spectroscopy mapping and sp³ hybridization** (a) Representative Raman spectra showing G, 2D, D, D′, and D+D′ peaks measured in a sample of graphene on an activated SiO$_2$/Si substrate (blue curve, sample B) and a pristine graphene sample (orange curve). (b) The scatter plot shows the intensity ratio $I_D/I_G$ versus $I_{D'}/I_G$. The red line fits to the data with a slope of 13.6. (c) The intensity ratio, $I_D/I_G$, as a function of the G peak positions ($\omega_G$).

### 3.3 Raman spectroscopy and sp³ hybridization.

Thus far, the transport data indicate the presence of energy gap formation in the samples of monolayer graphene on activated SiO₂/Si substrates. The Raman spectroscopy of the graphene devices provides further insight into the occurrence of the energy gap. A representative Raman spectrum of sample B is shown in Figure 3a and is compared to that of pristine graphene. Both Raman spectra show two characteristic peaks, the G peak at 1580 cm⁻¹ and 2D peak at 2670 cm⁻¹. The 2D peak, caused from the second order vibration in crystalline graphene,[36] is more apparent for pristine graphene. Conversely, the D (~ 1340 cm⁻¹) and D′ (~ 1610 cm⁻¹) peaks were more prominent and the 2D peak was smaller in sample B. Because the D band is associated with a double-resonance defect-mediated process,[37, 38] the appearance of D band peaks indicates the presence of defects that alter the original sp² graphitic network in crystalline graphene.

The D and D′ peaks of the Raman spectra are attributed to the defects in these graphene samples; these defects can be classified as the sp³ hybridization of C atoms,[29, 39] vacancies,[40, 41], or grain boundary.[42] The intensities of the D and D′ peaks indicate the density of the defects regardless of their origin. However, it has been shown that the intensity ratio $I_D/I_{D'}$ can shed more light into the nature of the defects.[43] It was found that sp³–type defects exhibit the highest ratio of $I_D/I_{D'}$ (~13), while vacancy-like defects and grain boundaries are lower (~7 and ~3.5, respectively).[43] Figure 3b shows the intensity ratio of $I_D/I_G$ as a function of $I_{D'}/I_G$ extracted from Raman spectra mapping on sample B; the data points follow a relatively linear relation. We found that the slope of the linear fit is ~13.6, indicating that the defects are due to the $sp^3$ hybridization of C atoms (more details in Supporting Information S3).

Next, we discuss the implications of the $sp^3$ hybridization of C atoms on the semiconducting behaviors observed in functionalized graphene. When graphene samples are fully functionalized, a theoretical band gap of ~1–2 eV is predicted; this value is much greater than the observed $E_G$ (~100 meV). Hence, we infer that the graphene is only partially functionalized, introducing randomly scattered insulating regions caused by the functionalization.[29, 30] We then model our graphene sample as intrinsic graphene mixed with scattered insulating regions, therefore creating a network of graphene channels. The observed transport gap can then be understood as a result of carrier confinement[44, 45] or localized effects due to edge roughness.[46, 47] Based on this proposed model, we first estimate an average channel width using the following empirical equation:[40]

$$\frac{I_D}{I_G} = \frac{102}{L_D^2},$$

where $L_D$ is the average distance (in nm) between the centers of the insulating regions. We then take the average $I_D/I_G$ of the whole sample to be $I_D/I_G \sim 4$ (Figure 3c), which corresponds to $L_D \sim 6$ nm. Assuming a channel width of 6 nm, we can then deduce $E_G$ to be on the order of 100 meV based on theoretical calculations,[45, 48] which is comparable to the observed energy gap.

We now discuss possible scenarios for the occurrence of the $sp^3$ hybridization of C atoms, as revealed by the Raman spectroscopy analysis. Oxygen plasma treatment of the $SiO_2/Si$ substrates causes the $SiO_2$ surface to be dominated with silanol groups.[32, 49] When the $SiO_2/Si$ substrate is enriched with silanol groups, theoretical calculations show that chemisorption of graphene is energetically favored.[50] Therefore, the observed $sp^3$-hybridization could be caused by the formation of C–H, C–OH, or C–O–Si bonds, altering the electronic properties of graphene and leading to the observed transport gap. Moreover, water molecules are likely absorbed on the surface because of the hydrophilic nature of silanol groups on the $SiO_2$ surface and may play a role in the electrochemical reaction.

**3.4 Carrier transport in disordered system.**

The presentation of randomly scattered defects indicated by the Raman spectroscopy allows us to interpret the low $T$ transport by the electron hopping mechanism. Figure 4a shows the $\sigma - V_G$ curves of sample B at $T$ ranging from 3 K to 180 K. The conductivity at the CNP decreases with decreasing $T$ and becomes lower than the quantum conductance ($e^2/h \approx 38.7$ μS) at $T \sim 50$ K, thus entering the insulating regime (Supporting Information S4). The metal-insulator transition has been reported in graphene samples functionalized by various methods.[9] Figure 4b shows a semi-log plot of $\sigma$ as a function of $T^{-1}$ at different $V_G$. We observed two distinct $\sigma - T$ behaviors at higher and lower $T$ regimes, separated by a crossover temperature, $T^*$, at approximately 10 K. At $T > T^*$, the carrier transport exhibits activated behavior, which is fitted by $\rho \sim \exp[(T_0/T)]$ and may be attributed to the nearest-neighbor hopping mechanism. For $T < T^*$, $\rho$ deviates from the simple activation behavior and can be reasonably fitted using $\rho \sim \exp[(T_1/T)^{1/3}]$. We therefore attribute the transport behavior in this $T$ regime to the variable-range hopping (VRH) model in disordered two-dimensional (2D) systems.[51] Figure 4c shows the data-fitted $T_0$ and $T_1$ plotted against $V_G$. We note that both $T_0$ and $T_1$ decrease with increasing carrier density, which can be explained by the restoration of metallic properties as the Fermi level approaches the mobility edge.[52]

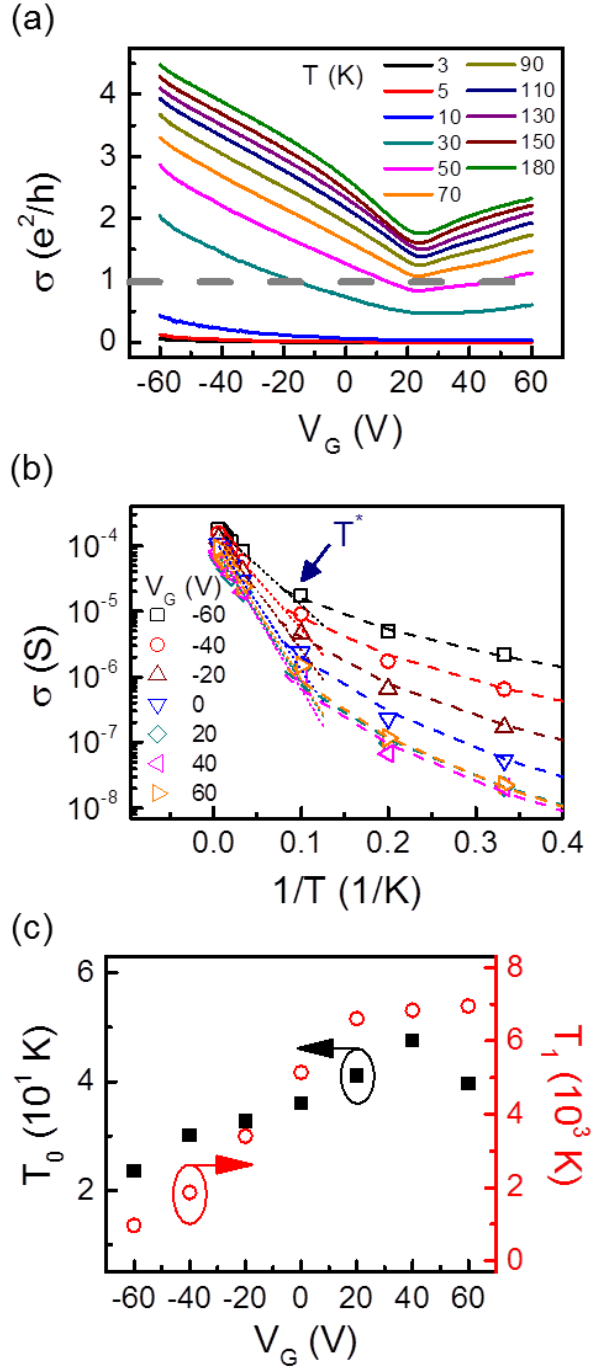

**Figure 4. Hopping transport behavior of a graphene on activated SiO$_2$/Si device** (a) $\sigma - V_G$ curves of sample B at $T$ ranging from 3 K to 180 K; the gray dashed line marks the quantum conductance, $e^2/h \approx 38.7$ μS. (b) The Arrhenius plot of $\sigma$ versus $T^{-1}$ at different $V_G$. Two distinct $\sigma - T$ behaviors can be identified at high and low $T$ regimes separated by a crossover

temperate at $T^* = 10$ K. The dotted and dashed lines are the results fitted to $\rho \sim \exp[(T_0/T)]$ and to $\rho \sim \exp[(T_1/T)^{1/3}]$ in the high and low $T$ regimes, respectively. (c) The characteristic temperatures $T_0$ and $T_1$ obtained from data fitting versus $V_G$ based on the thermal activation and the variable-range hopping mechanism.

Within the framework of the hopping mechanism, we further estimate the localization length ($\xi_{VRH}$)[53] based on the characteristic temperature $T_1$:

$$\xi_{VRH} = \sqrt{\frac{13.8}{k_B T_1 g}},$$

where $g$ is the density of states of graphene. Because the sample is in localized conduction regime, we use the carrier density as an approximation of $g$. We also assume that carrier density $n = C_G \sqrt{\Delta V_G^2 + V_0^2}/e$ to account for the fluctuation-induced electron-hole puddles near the Dirac region,[9] where $\Delta V_G$ is the difference between $V_G$ and the conduction minimum, and $V_0$ is chosen to be 30 V in accordance with the transport gap. We then obtain $\xi_{VRH} \approx 6-13$ nm, with the minimum hopping length near the vicinity of the CNP. As a comparison, we estimate the 2D localization length from the scaling theory:[54]

$$\xi_{2D} \sim \ell \exp\left(\frac{\sigma}{e^2/h}\right),$$

where $\ell = 2e^2 k_F v_F \sigma/h$ is the mean free path in semi-classical diffusion theory. We deduce that $\xi_{2D} \approx 10$ nm in the vicinity of the CNP at 50 K, which is comparable to $\xi_{VRH}$, indicating the validity of hopping transport through localized states at low temperatures. Notably, these localization lengths agreed well with the calculated average distance between the defects ($L_D$) from the Raman spectroscopy analysis, suggesting a correlation between the carrier localization and sp$^3$-type defects. Now, we can further analyze the crossover temperature, $T^*$, which is estimated using $T_0$ as: $T^* = [1/\pi(a/\xi)]T_0$, where $a$ is the optimized distance between two hopping sites and $\xi$ is the localization length. We assume $a \sim L_D$ from its relationship to the functionalization, and we evaluate $\xi$ from the scaling theory. Our analysis yields $T^* \approx 0.189 T_0 \sim 7.78$ K, which agrees reasonably well with the observed $T^* = 10$ K.

### 3.5 Magnetotransport in a disordered system.

Finally, we present the magnetotransport measurements of the functionalized graphene to further investigate disorder within the graphene samples. Figure 5a shows the temperature dependence of the conductance versus magnetic field ($\sigma - B$) curves of sample B at the CNP ($V_G = 23$ V) with $T$ ranging from 5 K to 180 K. The magnetoresistance (MR) in graphene can be mainly attributed to weak localization,[55] electron-electron interaction (EEI),[56, 57] and formation of charge puddles.[58] The MR caused by EEI is manifested by parabolic curves [56, 57] which was not observed in our graphene samples. Moreover, MR due to EEI was reported in relatively large and homogeneous samples, which are very different from the mechanically exfoliated and functionalized graphene samples used in this study. The formation of electron-hole puddles can account for low-field MR at low temperature.[58] However, we aim to discuss MR at magnetic field ranging from 0 to 6 T and temperature ranging from 5 to 160 K. We found that most MR data can be attributed to weak localization in graphene in these broad experimental parameters.[55] Moreover, because the conduction is dominated by carrier localization, weak localization is a reasonable scheme to be considered. At $70 < T < 160$ K, the magnetoconductivity (MC) began to decline at a magnetic field of approximately $4 - 5$ T. It is known that WL is suppressed in high crystalline graphene[55] and can be restored when inter-valley scattering due to short-range defects occurs.[59, 60] Therefore, the pronounced WL in the graphene samples suggests the presence of atomically sharp defects, which is consistent with the aforementioned $sp^3$–type defects. At higher temperatures ($T > 160$ K), the electron-phonon scattering increases, leading to quenching of the quantum interference and WL.

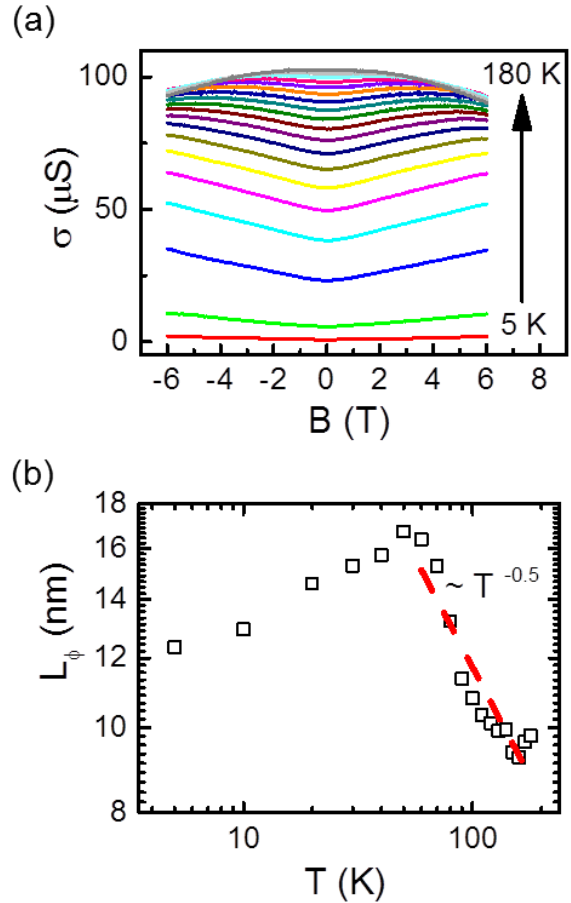

**Figure 5. Magnetotransport measurement and weak localization** (a) $T$ dependence of the $\sigma - B$ curves for sample B at the CNP ($V_G = 23$ V) ranging from 5 K to 180 K. A negative magnetoconductivity corresponding to weak localization is observed up to $T = 120$ K. (b) Phase coherence length ($L_\phi$) versus $T$ for sample B. The red dashed line indicates the linear temperature dependence of $\sim T^{-0.5}$ for $T > 50$ K.

We further analyzed the MC using the WL theory developed for graphene.[60, 61] The quantum correction to the semi-classical (Drude) conductivity, $\delta\sigma(B)$, is given by

$$\delta\sigma(B) = \frac{e^2}{\pi h}\left[F\left(\frac{\tau_B^{-1}}{\tau_\phi^{-1}}\right) - F\left(\frac{\tau_B^{-1}}{\tau_\phi^{-1} + 2\tau_i^{-1}}\right) - 2F\left(\frac{\tau_B^{-1}}{\tau_\phi^{-1} + \tau_i^{-1} + \tau_*^{-1}}\right)\right],$$

$$F(z) = \ln z + \psi(0.5 + z^{-1}),$$

$$\tau_B^{-1} = 4eDB/\hbar,$$

$$\tau_{\phi,i,*} = L^2_{\phi,i,*}/D,$$

where $\psi(x)$ is the digamma function and $D$ is the diffusion constant. Good agreement between our data and the WL theory was found for the entire $T$ range (Supporting Information S5). By fitting the $\sigma - B$ curves at different $T$ values, we obtained the phase coherence length $L_\phi$, as shown in Figure 5b. At $T > 50$ K, $L_\phi$ follows a $T$ dependence of $\sim T^{-0.5}$, which can be attributed to the phase randomization process in a metallic system.[55] However, at $T < 50$ K, $L_\phi$ in our sample was suppressed, which may be a result of a transition to strong localization (WL theory may not accurately describe the data in this regime).[9] It is noted that the $L_\phi$, which occurs at ~ 10 nm in our graphene samples, is much smaller than that of graphene with high crystallinity (several hundred nm),[61] indicating a high density of defects in our samples due to functionalization. Moreover, the maximum $L_\phi$ is comparable to the localization length and average distance between defect sites ($L_D$) discussed earlier.

## 4. Conclusion

In summary, we present a novel method for graphene functionalization by mechanically exfoliating graphene onto plasma-activated substrate surfaces. Pronounced semiconducting transport behaviors including nonlinear transport characteristics and an insulating regime in differential conductance mapping, strongly suggest the presence of an energy gap in monolayer graphene samples. Raman spectroscopy analysis provides solid evidence of the sp$^3$ hybridization of C atoms in the functionalized graphene. Our study introduces a feasible process for achieving semiconducting graphene-based materials with the use of chemically activated substrate surfaces for large-scale electronic applications.

**Acknowledgements**

W. W. would like to thank Jim Jr-Min Lin, Mei-Yin Chou, Yuh-Lin Wang, and Jer-Lai Kuo for insightful discussions. This work was supported by the Ministry of Science and Technology of Taiwan under contract numbers MOST 103-2112-M-001-020-MY3.

**Appendix A. Supplementary Data.**

Additional experimental details including device fabrication, estimation of the transport gap, spatial Raman spectroscopy profiles, temperature dependent $I-V_{SD}$ curves, and weak localization analysis can be found in the online version.

Supplementary Data for

# Demonstration of distinct semiconducting transport characteristics of monolayer graphene functionalized via plasma activation of substrate surfaces

Po-Hsiang Wang, Fu-Yu Shih, Shao-Yu Chen, Alvin B. Hernandez, Po-Hsun Ho,

Lo-Yueh Chang, Chia-Hao Chen, Hsiang-Chih Chiu, Chun-Wei Chen,

and Wei-Hua Wang*

*Corresponding Author.

Tel: +886-2-2366-8208. E-mail: wwang@sinica.edu.tw (W.-H. Wang)

## S1. Device fabrication.

**Device fabrication.**

For the control samples, graphene flakes were mechanically exfoliated onto OTS-modified $SiO_2$/Si substrates, in which the OTS surface modification has been described elsewhere.[1] In short, $SiO_2$ (300 nm)/Si wafers chips were first sonicated sequentially in acetone, $H_2O$:$NH_4OH$:$H_2O_2$ (50:1:1), and isopropanol solutions, followed by a brief (65W, 1 min) oxygen plasma cleaning before OTS treatment in a glovebox. Graphene devices described in this report were subjected to a different substrate treatment procedure. Prior to graphene flake deposition, the $SiO_2$/Si substrates (heavily n-type doped silicon wafers capped with 300 nm $SiO_2$) were sonicated in acetone, isopropanol alcohol, and deionized (DI) water baths, in that order. The cleaned substrates were baked on a hotplate at 100 °C before plasma treatment. Oxygen plasma was generated by a homemade plasma etcher with a base pressure of oxygen at 450 mTorr and a flow rate of ~ 50 sccm and a radiofrequency (RF) power of 10 – 50 W. After 10 min of plasma treatment, the activated substrates were immediately dipped in DI water, followed by blow-drying with $N_2$. Monolayer graphene was mechanically exfoliated onto the pre-processed substrates. The graphene flakes were identified by optical microscopy and examined by Raman spectroscopy.

A flowchart of device fabrication procedure was presented in Figure S1a. Figure S1b shows a schematic diagram of a graphene device and a corresponding optical image of a typical graphene device is shown in Figure S1c.

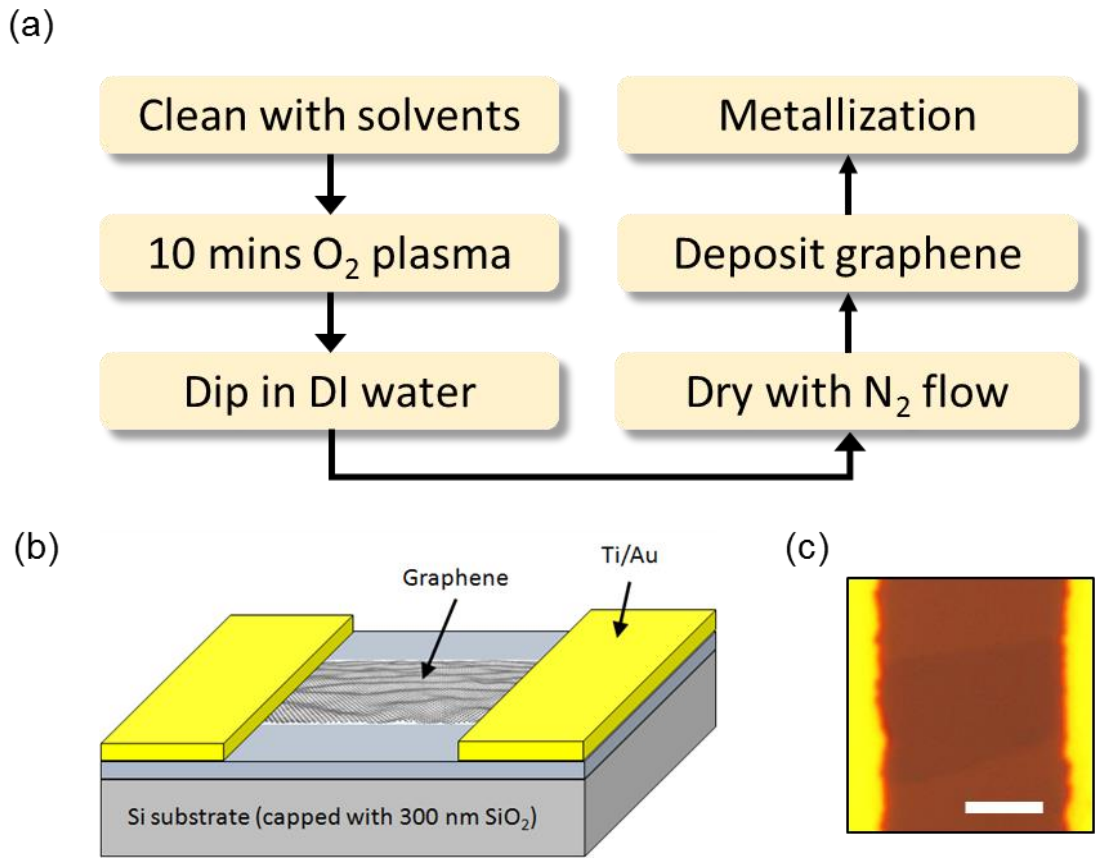

**Figure S1.** (a) A flowchart of the procedure for device fabrication of graphene on activated SiO$_2$/Si substrates. (b) A schematic diagram of a graphene device on an activated SiO$_2$/Si substrate. (c) An optical image of a typical graphene device is shown. The scale bar is 5 μm.

We note that the condition for oxygen plasma treatment was very different from the general procedures, e.g. Nagashio *et al*. [2]. We emphasize two key differences: (1) the plasma treatment duration of 10 min is much longer than the general procedure, e.g. duration ~ 10 sec used in previous report [2]; (2) The base pressure of oxygen (450 mTorr) is lower than what we commonly used (~ 800 mTorr) for cleaning recipe. We aim for a less uniform reaction on the surface and our method results in greater surface roughness, as evidenced by the AFM characterization.

Figures S2 compares the topography of $SiO_2$/Si substrates undergone different plasma condition. It can be seen that the RMS roughness greatly increased from 0.17 nm (not treated) to 1.45 nm after the oxygen plasma treatment (10 W, 10 min). For comparison, the RMS roughness only increased to 0.27 nm by employing the conditions for cleaning recipe (60 W, 10 sec). We infer that surface roughness can be a critical factor in determining the chemical activation. After mechanical exfoliation, graphene tends to conform to the surface profile of the substrates, which results in graphene ripples. It is well known that graphene with larger curvature is subjected to higher chemical reactivity, and $sp^3$ hybridization may occur at the curved area of graphene.[3]

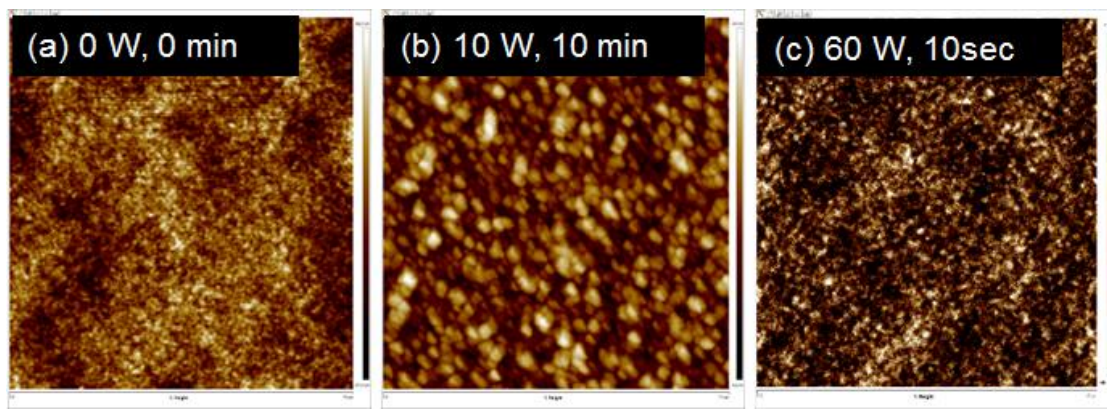

**Figure S2**. Surface topography obtained by AFM characterization for (a) pristine SiO$_2$/Si substrates, (b) after the oxygen plasma treatment (10 W, 10 min), and (c) after the oxygen plasma treatment for cleaning procedure (60 W, 10 sec).

**Raman Spectroscopy and Electrical Measurements.**

Raman spectra were recorded under an excitation of a 532 nm laser light that is focused on the sample by a 100X Olympus objective with a nominal spot size of 1 μm. Scattered light was collected through the same objective and recorded by a spectrometer (iHR 550, Horiba). Electrical contacts of Ti/Au (5/50 nm) were deposited on the samples using electron-beam evaporation by employing TEM grids as shadow masks. Before the transport measurement, the samples were annealed *in situ* at 110 °C for 1-2 hours to remove adsorbates present from the ambient environment and the fabrication processes.

The Raman data of sample B shown in Figure 3a of the main text was taken after the transport measurement. The Raman spectra of sample D shown in Figure S5a of supporting information were taken before the transport measurement. In addition, Figure S3 shows Raman spectra of another device (sample E) which were also taken before the annealing process and the transport measurement. We therefore observed the evidence of sp$^3$ hybridization before the annealing process and transport measurement.

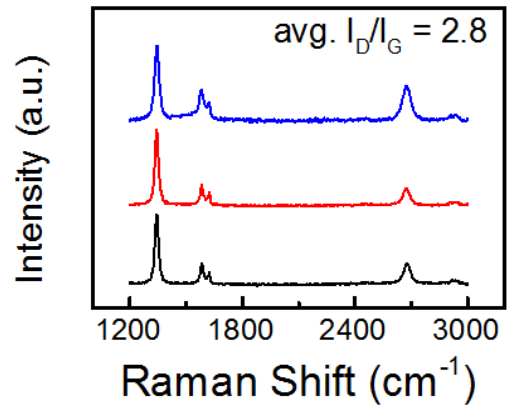

**Figure S3**. Raman spectra of sample E which exhibited pronounced D band. The spectra were taken before the annealing process and the transport measurement.

## S2. Estimation of the transport gap

We employ an energy gap estimation method different from the one in the main text (also described in the next paragraph). To determine the size of the energy gap we examine $dI/dV_{SD}$ versus $V_{SD}$ curve of sample A in the main text ($V_G = 60$ V), as shown in Figure S4a. Here, the nonlinear gap, $\Delta V_{SD}$, is defined by the value of $V_{SD}$ when $dI/dV_{SD}$ starts to increase. We utilize the first derivative of $dI/dV_{SD}$, shown as red solid line, to define this threshold (marked with black arrows), which corresponds to the value of $V_{SD}$ when the source and drain levels overlap with the band edges. We can then estimate the nonlinear gap, $\Delta V_{SD}$, as ~ 90 meV.

Here, we show how $\Delta V_{SD}$ is determined in Figure 4c of the main text. In Figure S4b, we plot the differential conductance traces as a function of $V_G$ for sample A at $V_{SD} = 0$ mV (black squares). A clear onset of differential conductance at ~ $10^{-8}$ S is identified. We then calculate the first derivative of the differential conductance data with respect to $V_G$ (the red curve); this can also be used to assign the off-current level as $2 \times 10^{-8}$ S and thus determine the color level. The energy gap is then obtained from the value of $V_{SD}$ at the vertices of the diamond-shaped area in the $dI/dV_{SD}$ mapping; this value is $E_G \sim 100$ meV for sample A. The differential conductance mappings as a function of $V_G$ and $V_{SD}$ for sample B in the main text and a third sample (sample C) are shown in Figure S4c and S4e, respectively. The estimated energy gaps for sample B and C are 100 and 80 meV, respectively.

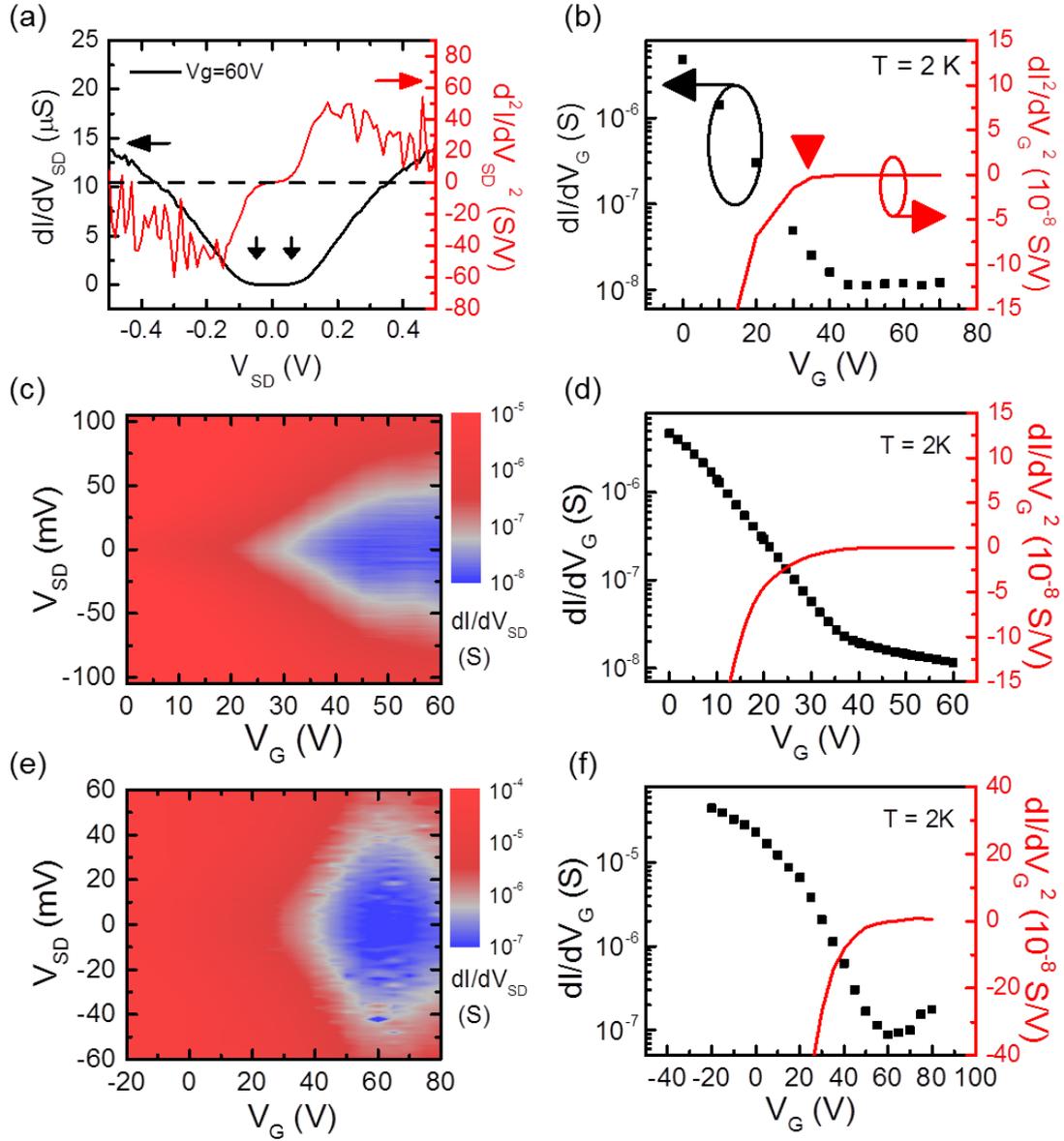

**Figure S4.** Energy gap estimation of graphene on activated SiO$_2$/Si substrates. (a) Differential conductance traces as a function of $V_{SD}$ for sample A at $V_G = 60$ V. (b) Differential conductance traces as a function of $V_G$ for sample A at $V_{SD} = 0$ mV. Red lines are the first derivatives of the black dots. Energy gap estimation using

differential conductance mapping for sample B (c) and sample C (e). Differential conductance traces as a function of $V_G$ for sample B (d) and sample C (f). Red lines are the first derivatives of the black dots.

## S3. Spatial Raman spectroscopy profiles.

Figure S5a shows the Raman spectra measured along the graphene channel of a fourth sample of graphene on activated SiO$_2$/Si substrate (sample D), in which a D peak is observed in all spectra. The Raman profiles of D, D′, G, and 2D peaks (Figure 4a in the main text and Figure S5a) were fitted to Lorentzian line shapes:

$$H(x) = H_0 + \frac{I}{\pi} \frac{(w/2)}{(x-x_c)^2 + (w/2)^2},$$

where $H_o$ is the baseline correction, $x_c$ is the position of the maximum, $w$ is the full width at half maximum (FWHM), and $I$ is the integral intensity. Further analysis of the spectra in Figure S5a using this fitting scheme shows that the average ratio of $I_D/I_G$ to $I_{D'}/I_G$ is approximately 13.6 (Figure S5b). A comparable ratio was obtained by analyzing the Raman mapping spectra of sample B, as discussed in the main text, and it was attributed to the sp$^3$ hybridization of C atoms. Figure S5c shows the intensity ratio, $I_D/I_G$, extracted from Figure S5a plotted as a function of the G band position ($\omega_G$). The intensity ratio ($I_D/I_G \approx 2.4$) of sample D is smaller than that of sample B indicating a lower degree of functionalization. In this study, a total of 20 graphene samples exhibited $I_D/I_G$ larger than 0.5. However, there was variation in the degree of functionalization, which is still not well understood, and precise control of the functionalization requires further study.

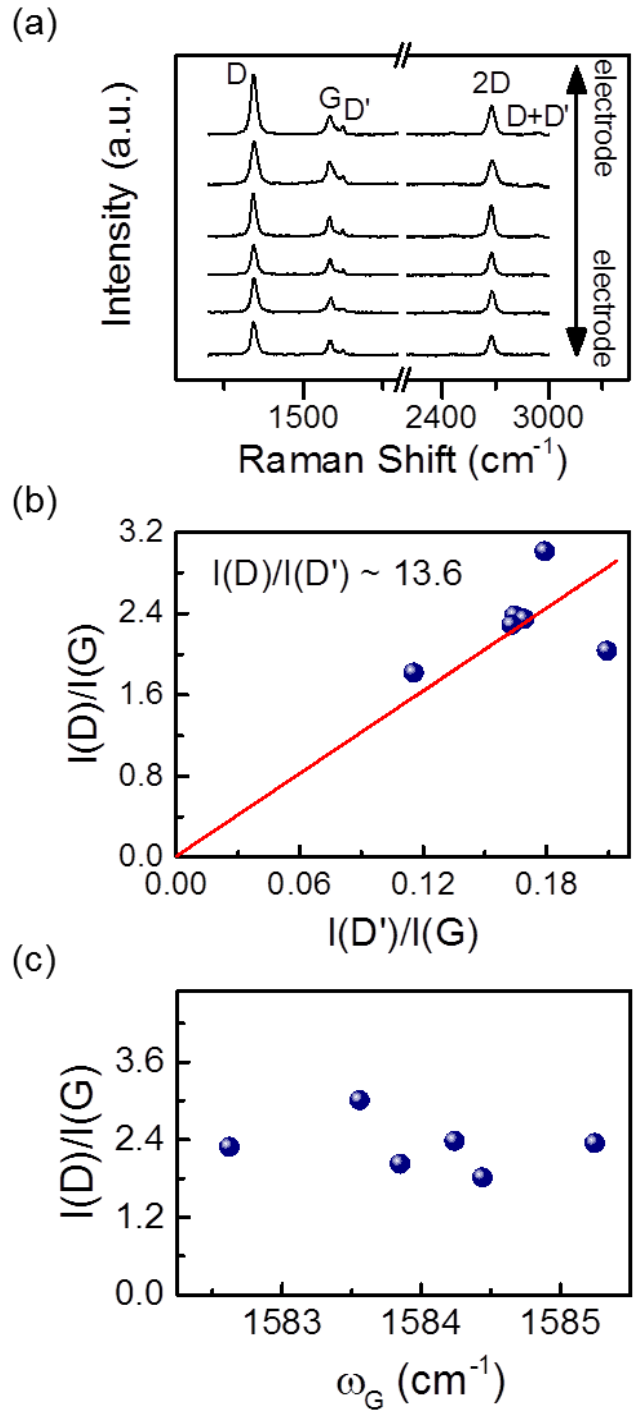

**Figure S5.** (a) Raman spectra of sample D measured along the graphene channel. (b) The intensity ratio, $I_D/I_G$, as a function of $I_{D'}/I_G$ corresponding to the spectra in (a). (c) The ratio $I_D/I_G$ as a function of the G band position ($\omega_G$).

## S4. Temperature dependent $I-V_{SD}$ curves

Figure S6 shows $T$ dependent $I-V_{SD}$ curves of sample B for a small DC source-drain bias voltage with $V_G$ fixed at CNP ($V_G = 23$ V). The large variation in device resistance is clearly observed as the nonlinearity of the $I-V_{SD}$ curve varies within the measured $T$ range from 2 K to 180 K. The graphene device changes from exhibiting metallic behavior (characterized with linear $I-V_{SD}$ curves with resistance of ~ 10 k$\Omega$) to insulating behavior (characterized with nonlinear $I-V_{SD}$ curves with resistance of several tens of M$\Omega$).

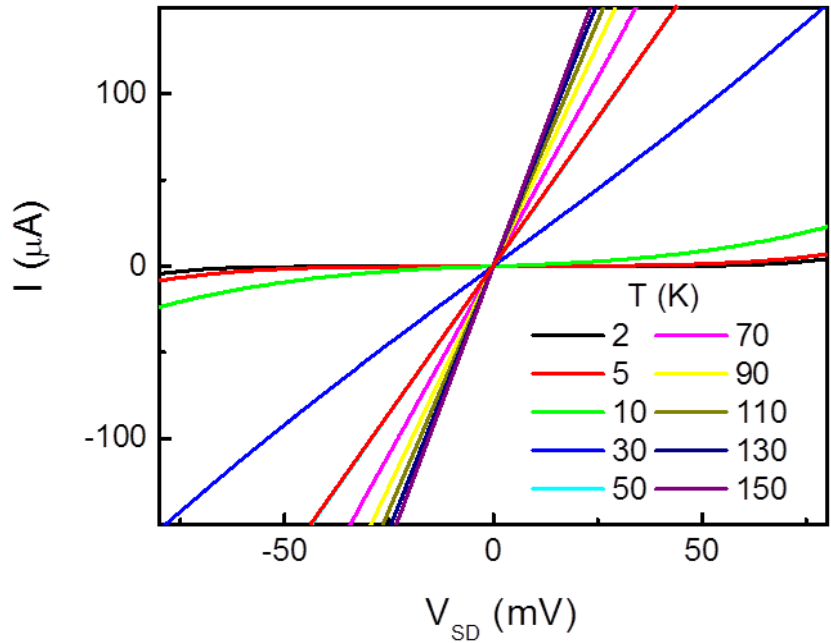

**Figure S6.** The $T$ dependence of $I-V_{SD}$ curves for sample B of graphene on an activated SiO$_2$/Si substrate. The graphene device transitions from exhibiting metallic behavior to exhibiting insulating behavior at approximately $T = 30$ K.

## S5. Weak localization.

Figure S7 illustrates two representative curves of our data at $T = 10$ and $T = 150$ K fitted to the WL theory. All experimental data agreed satisfactorily with the WL theory at all temperatures with an unambiguously determined $L_\phi$. On the other hand, the obtained scattering rates of both $\tau_i^{-1}$ and $\tau_*^{-1}$ showed a larger variation. Nevertheless, in the framework of semi-classical diffusive conductivity, both inter- and intra-valley scattering rates are correlated in determining the diffusion constant: $D = 1/2 \cdot v_F^2 \left( \tau_i^{-1} + \tau_*^{-1} \right)^{-1}$. Therefore, the uncertainty of $\tau_i^{-1}$ and $\tau_*^{-1}$ in the fitting process does not affect the calculation of phase coherence length discussed in Figure 5b of the main text.

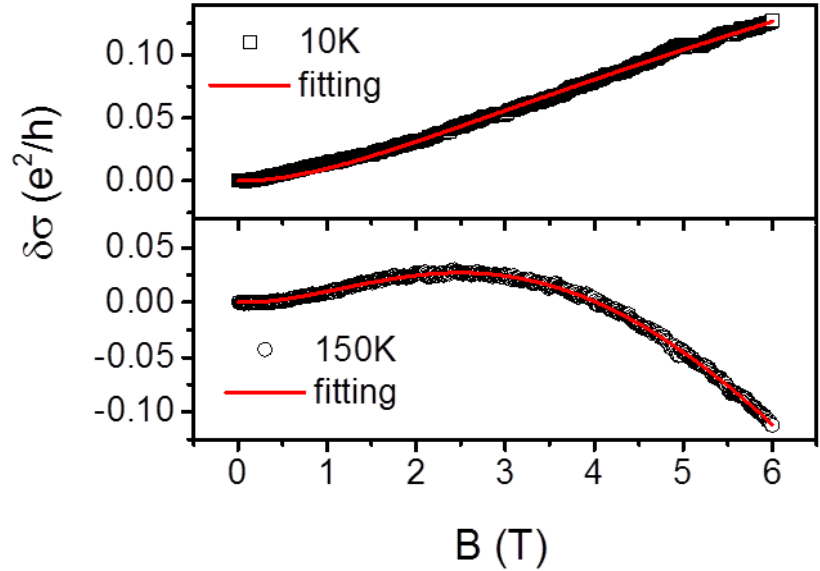

**Figure S7.** The representative experimental data ($T = 10$ and $T = 150$ K) fitted to WL theory. Black open squares correspond to experimental results while red lines correspond to theoretical trend.